\documentclass[usenatbib,fleqn]{mnras}

\usepackage{newtxtext,newtxmath}

\usepackage[T1]{fontenc}
\usepackage{ae,aecompl}

\usepackage{graphicx}	
\usepackage{amsmath}	
\usepackage{amssymb}	
\usepackage{ulem}

\usepackage{multirow}
\usepackage{array}  
\usepackage{booktabs}
\usepackage{times}
\usepackage{subfigure}

\usepackage[usenames,dvipsnames]{xcolor}

\usepackage[]{xcolor}

\definecolor{tbc}{cmyk}{0.05002,1,0.9,0}

\definecolor{mygreen}{cmyk}{0.85002,0.3,1,0}

\definecolor{myyellow}{cmyk}{0.00,0.24,0.86,0.20}

\definecolor{mypink}{cmyk}{0.00,1.00,0.00,0.00}

\newcommand{\obj}{J0946+1017}
\newcommand{\weimi}{$\mu$m}

\newcommand{\hb}{H$\beta$}
\newcommand{\mgii}{Mg\,{\sc ii}}
\newcommand{\mgiidoublet}{Mg\,{\sc ii}$\lambda\lambda2796,2803$}

\newcommand{\oiii}{[O{\sc iii}]$\lambda\lambda4959,5007$}

\newcommand{\nii}{[N{\sc ii}]$\lambda\lambda6548,6583$}

\newcommand{\feii}{Fe{\sc\,ii}}

\newcommand{\kmps}{$\rm km\,s^{-1}$}
\newcommand{\flux}{$\rm erg\,s^{-1}cm^{-2}$}
\newcommand{\lum}{$\rm erg\,s^{-1}$}

\newcommand{\sdss}{\mbox{SDSS}}
\newcommand{\boss}{\mbox{BOSS}}
\newcommand{\ts}{\mbox{TripleSpec}}
\newcommand{\wise}{\mbox{\it WISE}}
\newcommand{\rosat}{\mbox{\it ROSAT}}
\newcommand{\rass}{\mbox{\it RASS}}
\newcommand{\xmm}{\mbox{\it XMM-Newton}}
\newcommand{\fermi}{\mbox{{\it Fermi}/LAT}}

\title[SDSS~J0946+1017: a $\gamma$-ray emitting NLS1 galaxy]{
SDSS~J094635.06+101706.1: a redshift one, very radio-loud, $\gamma$-ray emitting narrow-line Seyfert 1 galaxy
}

\author[S. Yao et al.]{
Su Yao$^{1,2}$\thanks{E-mail: yaosu@pku.edu.cn}, 
S. Komossa$^{3}$, 
Wen-Juan Liu$^{4,5}$, 
Weimin Yi$^{4,5,6}$, 
Weimin Yuan$^{7}$, 
\newauthor 
Hongyan Zhou$^{8,9}$, Xue-Bing Wu$^{1,10}$
\\
$^{1}$Kavli Institute for Astronomy and Astrophysics, Peking University, Beijing 100871, China\\
$^{2}$National Astronomical Observatories, Chinese Academy of Sciences, Beijing 100012, China\\
$^{3}$Max-Planck Institut f{\"u}r Radioastronomie, Auf dem H{\"u}gel 69, 53121 Bonn, Germany\\
$^{4}$Yunnan Observatories, Chinese Academy of Sciences, Kunming, Yunnan 650011, China\\
$^{5}$Key Laboratory for the Structure and Evolution of Celestial Objects, Chinese Academy of Sciences, Kunming 650011, China\\
$^{6}$Department of Astronomy \& Astrophysics, The Pennsylvania State University, 525 Davey Lab, University Park, PA 16802, USA\\
$^{7}$Key Laboratory of Space Astronomy and Technology, National Astronomical Observatories, 
Chinese Academy of Sciences, Beijing 100012, China\\
$^{8}$CAS Key Laboratory for Research in Galaxies and Cosmology, Department of Astronomy, USTC, Hefei, Anhui 230026, China\\
$^{9}$SOA Key Laboratory for Polar Science, Polar Research Institute of China, 451 Jinqiao Road, Shanghai 200136, China\\
$^{10}$Department of Astronomy, School of Physics, Peking University, Beijing 100871, China\\
}

\date{Accepted XXX. Received YYY; in original form ZZZ}

\pubyear{2019}

\begin{document}
\label{firstpage}
\pagerange{\pageref{firstpage}--\pageref{lastpage}}
\maketitle

\begin{abstract}
As hybrids of narrow-line Seyfert 1 (NLS1) galaxies and blazars, 
$\gamma$-ray emitting NLS1s are important probes of jet physics in the high Eddington-ratio regime.
Only very few of them are known to date; the majority of them below redshift $z=0.5$. 
Here we present the identification of the $\gamma$-ray emitting 
AGN TXS\,0943+105 (SDSS~J094635.06+101706.1) as a high-redshift NLS1 galaxy. 
It turns out to be
one of the radio-loudest NLS1s known, 
highly variable at all wavelengths, 
and shows widely extended radio emission at a (projected) $>$ 100 kpc scale. 
It is a known strong $\gamma$-ray emitter with a luminous flare reported previously. 
At redshift $z$=1.004, this is the most distant $\gamma$-NLS1 known to date. 
\end{abstract}

\begin{keywords}
galaxies: active -- galaxies: nuclei -- galaxies: Seyfert -- accretion, accretion discs -- galaxies: jets 
\end{keywords}

\section{Introduction}
\label{intro}

Radio-loud and $\gamma$-ray emitting narrow-line Seyfert 1 (NLS1) galaxies are important new probes of the formation and evolution of radio jets, 
in a regime different from classical blazars \citep[see a review by][]{2018rnls.confE..15K}. 
NLS1s are a subgroup of AGN with extreme properties in the
optical and X-ray band, and therefore stand out in AGN correlation space 
\citep[e.g.][]{1992ApJS...80..109B, 2010ApJS..187...64G, 2012AJ....143...83X}. They harbor supermassive black holes (SMBHs) of
relatively low masses, accreting at a high
rate, and therefore represent a class of AGN which are rapidly growing their SMBHs in the local universe. 

While only studied in the optical and X-ray regime for some decades
and preferentially radio-quiet, 
nevertheless a population of radio-loud, jet-emitting NLS1
galaxies was identified \citep{2006AJ....132..531K, 2008ApJ...685..801Y}. 
These combine the properties of blazars on the one hand, and NLS1 galaxies on the other hand. 
With \fermi, it was discovered that a few of the jetted NLS1 galaxies emit luminous $\gamma$-ray radiation, highly variable in some cases, 
and independently confirming the presence of powerful jets in some NLS1 galaxies \citep[][]{2009ApJ...707L.142A}. 
Meanwhile, about 15 $\gamma$-ray emitting NLS1s
have been identified 
(\citealt{2009ApJ...699..976A};
\citealt{2009ApJ...699..976A};
\citealt{2011nlsg.confE..24F};
\citealt{2012MNRAS.426..317D};
\citealt{2015MNRAS.452..520D};
\citealt{2015MNRAS.454L..16Y};
\citealt{2018MNRAS.477.5127Y};
\citealt{2018ApJ...853L...2P};
\citealt{2018A&A...614L...1L}; 
see \citealt{2018rnls.confE..15K} and 
\citealt{2018MNRAS.481.5046R} for reviews).

NLS1 galaxies as a group are spectroscopically defined by the small widths of their broad \hb\ emission lines, 
with full width at half maximum (FWHM) typically taken to be less than 2000\,\kmps, 
strong \feii\ emission complexes and weak forbidden lines 
\citep[][]{1985ApJ...297..166O}. 
Most NLS1s
have been reported at $z\lesssim0.8$, 
including the ones detected at $\gamma$-rays, 
as their identification requires \hb\ falling into the spectral coverage of large surveys. 
An exception is SDSS\,J1222+0413 at $z$=0.966 \citep[][]{2015MNRAS.454L..16Y}. 

Here, we report the identification of  TXS\,0943+105 (SDSS J094635.06+101706;
J0946+1017 hereafter) at redshift $z$=1.004 
as a new very radio-loud NLS1 galaxy 
which is also $\gamma$-ray detected \citep[][]{2015ApJS..218...23A}. 
The optical identification is based on the Sloan Digital Sky Survey-III Baryon Oscillation Spectroscopic Survey \citep[SDSS-BOSS,][]{2013AJ....145...10D},
which for the first time included the H$\beta$ regime (Section~\ref{sec:spectroscopy}) because of the broader wavelength range covered. 
Multi-wavelength follow-up observations are presented in Section~\ref{sec:multiband}, 
while Section~\ref{sec:discussion} provides the discussion and conclusions. 
Throughout this paper we use a $\Lambda$CDM cosmology with $H_0=70$ km s$^{-1}$ Mpc$^{-1}$, $\Omega_\Lambda=0.73$ and $\Omega_{\rm M}=0.27$.

\section{Spectroscopy}
\label{sec:spectroscopy}

\subsection{Observations}

The optical spectrum of \obj\ was obtained on January 27, 2012 (MJD 55953) 
during SDSS-III's Baryon Oscillation Spectroscopic Survey \citep[BOSS,][]{2013AJ....145...10D} by the Sloan 2.5m telescope 
equipped with a spectrograph which covers a wavelength range of 3600-10400\,\AA\ and has a spectral resolution of $R=1560$-$2650$. 
We retrieved the calibrated 1D spectrum from the SDSS archive. 

During the 2016A cycle of the Telescope Access Program of China, 
we obtained a near-infrared (NIR) spectrum of \obj\ using the TripleSpec spectrograph mounted on the Palomar 5\,m Hale telescope on April 20 (UT), 2016 (MJD 57498). 
The spectrograph provides a wavelength coverage from 0.9 to 2.46\,\weimi\ at a resolution of 1.4-2.9\,\AA\ with two gaps at approximately 1.35 and 1.85\,\weimi\ owing to telluric absorption bands \citep[][]{2004SPIE.5492.1295W}. 
For optimal removal of the strong sky background in the NIR, 
we observed target (A) and off-target (B) positions in an A-B-B-A dither pattern along the slit with four exposures of 300\,s. 
The sky was clear with seeing $\sim$1.2\arcsec and the slit-width was fixed to 1\arcsec. 
The data reduction, including flux calibration and telluric correction, 
was carried out with the Interactive Data Language (IDL)-based package {\tt SpexTool} \citep[][]{2004PASP..116..362C}, 
as described in detail in \citet{2015ApJ...799..189Z}.

\subsection{Data Analysis}
\label{sec:spec_fitting}

\begin{figure*}
	\centering
	\includegraphics[width=0.95\textwidth]{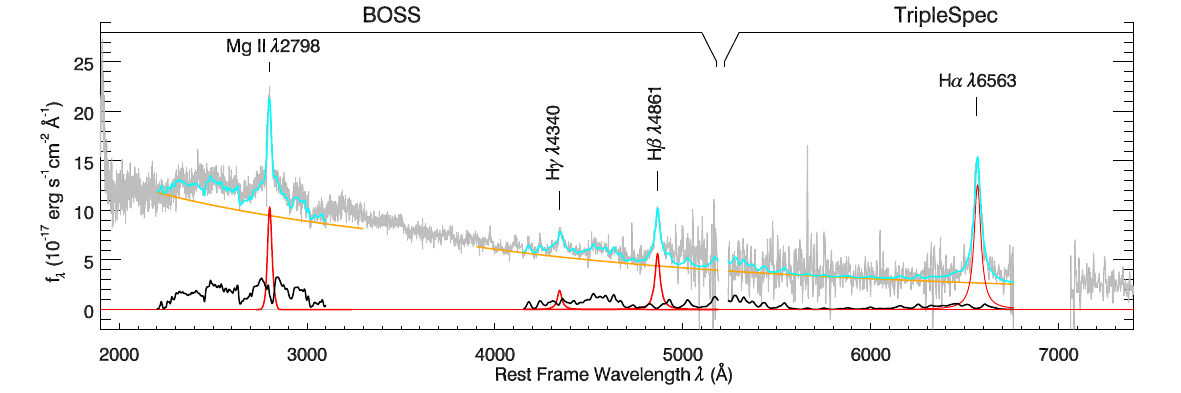}
	\caption{
	The combined rest-frame spectrum (grey) of \obj\ after correction for Galactic extinction. 
	The bottom black curves represent the best-fit \feii\ multiplets underneath the \mgii\ and Balmer lines. 
	The orange curves represent the best-fit single power law model and the red curves provide the best-fit broad emission line models. 
	The cyan curves represent the combination of the models. 
	\label{fig:spec}}
\end{figure*}

Before the spectral fitting, 
both \boss\ and \ts\ spectra are corrected for Galactic extinction with $E(B-V)=0.025$\,mag \citep[][]{2011ApJ...737..103S} and an $R_{V}=3.1$ extinction law, 
and then are transformed into the source rest frame at a redshift $z=1.004$ provided by the SDSS spectroscopic pipeline. 
The fitting is based on IDL routines in the {\tt MPFIT} package \citep[][]{2009ASPC..411..251M}, 
which performs a $\chi^{2}$ minimization using the Levenberg-Marquardt method. 
Due to flux variability of the source and non-simultaneity of the spectroscopy, 
the \sdss-\boss\ and TripleSec spectra had to be re-scaled. 
We fit the \boss\ spectrum using a single power law in the fitting windows which are thought to be free of any strong emission lines: 
[3020, 3030]\,\AA, 
[3790, 3810]\,\AA, 
[4200, 4210]\,\AA\  
and 
[5080, 5100]\,\AA\ 
\citep[][]{2001AJ....122..549V}. 
Then the \ts\ spectrum is scaled so that the spectrum in the range
[5300, 6400]\,\AA\ matches the best-fit single power law. 
The \boss\ spectrum and the scaled \ts\ spectrum are combined into one spectrum (Figure~\ref{fig:spec}). 
We use this combined spectrum in the following analysis.

In order to measure the primary emission lines, 
we adopt a similar procedure as in \citet{2015MNRAS.454L..16Y}, 
which models the emission lines and their underlying continuum simultaneously. 
We perform spectral fitting separately for the Balmer lines in the optical and the \mgii\ line in the ultraviolet. 
During the fitting procedure, 
a pseudo-continuum consisting of a simple power law and \feii\ multiplets is used. 
The optical and ultraviolet \feii\ are modelled using templates provided in \citet{2004A&A...417..515V} and \citet{2006ApJ...650...57T}, respectively.

The broad Balmer lines including H$\alpha$ were modelled with either a Lorentzian or a concentric double-Gaussian profile. 
Narrow lines potentially present in the spectrum, including possible narrow Balmer lines, \oiii\ and \nii\ are described by a single Gaussian profile with $\rm FWHM\leqslant900$\,\kmps. 
The broad Balmer lines are constrained to have the same width and redshift, 
and a similar constraint is set for all the narrow lines. 
Then the Balmer emission lines are fitted simultaneously with a pseudo-continuum in the range of [4160, 5100]\,\AA\ and [6000, 6800]\,\AA.

To measure the \mgii\ line, 
line profile and pseudo-continuum are fitted simultaneously to the spectrum in the range of 
[2200, 3100]\,\AA. 
We fit the \mgiidoublet\ doublet using a similar model as in \citet{2009ApJ...707.1334W}. 
Each line of the doublet is modeled with one broad and one narrow component. 
The broad component is a truncated five-parameter Gauss-Hermite profile \citep[e.g.][]{1993ApJ...407..525V}. 
Both broad components are assumed to have the same width and redshift, 
and their flux ratio is constrained between 2:1 and 1:1 \citep[][]{1997ApJ...489..656L}. 
The narrow component is a single Gaussian profile with constraint of $\rm FWHM\leqslant900$\,\kmps. 
The redshift and line flux ratio of the two narrow components are constrained following the same prescription as for the broad  components. 
We find, however, that addition of narrow components in Mg II is not needed.

No other narrow lines are detected in the spectrum, either. 
The Balmer lines can be similarly well fitted with either a Lorentzian or a double Gaussian profile, 
with $\rm FWHM(H\beta_{broad}) = 1974\pm51$\,\kmps\ for a Lorentzian and $2199\pm30$\,\kmps\ for a double-Gaussian profile, respectively. 
These measurements imply a NLS1 classification of \obj. 
This classification is independently confirmed by the faintness of [O{\sc~iii}]
as well as the very strong \feii\ emission complexes,  
with $R_{4570}\equiv$Fe\,{\scshape ii}\,$\lambda4570$/H$\beta_{\rm total}\approx1.0$, 
where Fe{\scshape\,ii}\,$\lambda4570$ is the flux of \feii\ integrated from 4434\,\AA\, to 4684\,\AA\ 
\citep[e.g.][]{2001A&A...372..730V}. 
The fitting results are summarized in Table~\ref{tab:spec_fitting}.

\begin{table*}
	\caption{Fit results for the main emission lines. }
	\label{tab:spec_fitting}
	\begin{center}
	\setlength{\tabcolsep}{3.0pt} 
	\begin{tabular}{ccccccccccccccc}
		\hline
		& $\rm FWHM(H\beta_{broad})$ & $f(\rm H\alpha_{broad})$ & 
			$f(\rm H\beta_{broad})$  &  $f(\rm H\gamma_{broad})$ & 
			$R_{4570}$ & 
			FWHM(\mgii$_{\rm broad})$ & $f$(\mgii$_{\rm broad})$ \\
		& [\kmps] & [$\rm erg\,s^{-1}\,cm^{-2}$] & 
			[$\rm erg\,s^{-1}\,cm^{-2}$] & [$\rm erg\,s^{-1}\,cm^{-2}$] & 
			& 
			[\kmps] & [$\rm erg\,s^{-1}\,cm^{-2}$]  \\
		\hline
		Lorentzian & 1974$\pm$51 & $(17.24\pm0.42)\times10^{-15}$ & $(5.76\pm0.20)\times10^{-15}$ & 
			$(1.75\pm0.11)\times10^{-15}$ & 0.90$\pm$0.08 & 
			\multirow{2}{*}{2919$\pm$196} & \multirow{2}{*}{$(6.52\pm0.21)\times10^{-15}$} \\ 
		Double-Gaussian & 2199$\pm$30 & $(17.58\pm1.02)\times10^{-15}$ & $(5.44\pm0.41)\times10^{-15}$ & 
			$(2.32\pm0.06)\times10^{-15}$ & 
			1.04$\pm$0.04 & \\
		\hline
	\end{tabular}
	\parbox[]{\textwidth}{
	{\it Notes.} FWHMs are not corrected for instrumental resolution. 
	}
	\end{center}
\end{table*}

\section{Multi-wavelength Properties}
\label{sec:multiband}

\subsection{Radio emission}
\label{sec:radio}

\obj\ was detected by several radio surveys. 
Its highest precision radio coordinate from Very Long Baseline Radio Interferometry observations \citep[VLBI, RA=09h46m35.1s, Del=+10\degr17\arcmin06\arcsec.1,][]{2015AJ....150...58F} agrees within better than 0.1 arcsec with its Gaia \citep[RA=09h46m35.1s, Del=+10\degr17\arcmin06\arcsec.1,][]{2018A&A...616A...8A} and SDSS coordinates.
The Texas Survey measured a flux density of $505\pm28\rm\,mJy$ at 365\,MHz 
with a spectral index of $\alpha_{\rm rad}\sim-0.5$ ($S_{\nu}\propto\nu^{\alpha}$) estimated between 335-380\,MHz, 
but the error in this spectral index is naturally high  
given the small frequency range involved 
\citep[][]{1996AJ....111.1945D}. 
The non-simultaneous observations by the Green Bank telescope imply flux densities of $379\rm\,mJy$ at 1.4\,GHz and $294\rm\,mJy$ at 4.85\,GHz, respectively, 
and the spectral index between 1.4 and 4.85\,GHz is flat with $\alpha_{\rm rad}\sim-0.2$ 
\citep[][]{1991ApJS...75....1B, 1992ApJS...79..331W}. 
In the higher frequency range, 
\obj\ was observed at 8.4\,GHz with the Very Large Array (VLA) within the framework of the Combined Radio All-Sky Targeted Eight GHz Survey \citep[CRATES,][]{2007ApJS..171...61H} with $\sim293$\,mJy flux density and at 22\,GHz with the VLBI Exploration of Radio Astrometry (VERA) instrument with $\sim90$\,mJy flux density \citep[][]{2007AJ....133.2487P}.

\begin{figure}
	\centering
	\includegraphics[width=0.70\columnwidth]{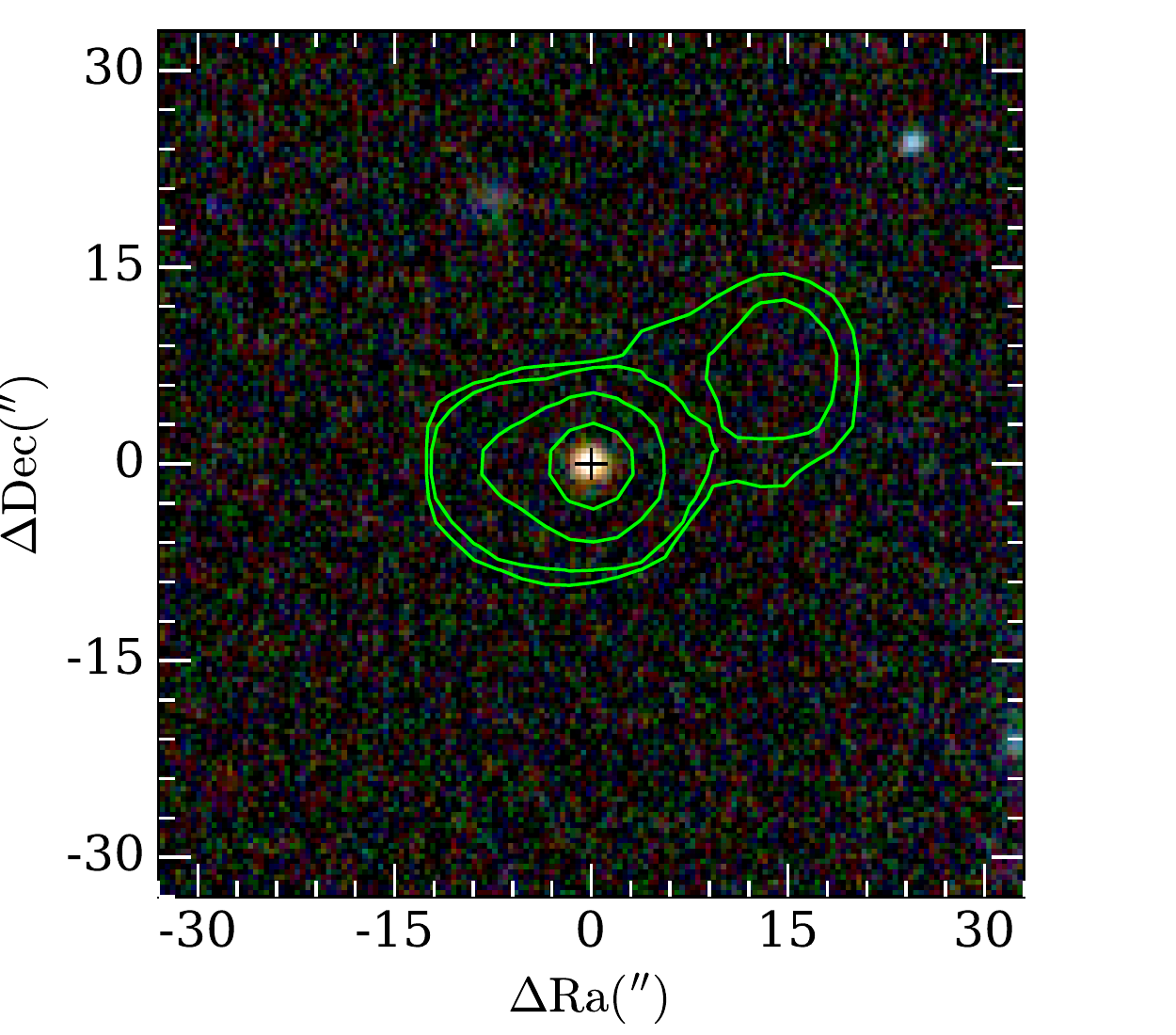}
	\caption{
	Pseudo-color RGB image created from SDSS $i$-$r$-$g$ band images centered at \obj's position (black `+'), 
	overlaid with the FIRST 1.4\,GHz image (green contours). 
	The contours are plotted as $(1,\,2,22,125)\times6\sigma$. 
	\label{fig:opt_radio}}
\end{figure}

The Faint Images of the Radio Sky at Twenty-Centimeters \citep[FIRST,][]{1995ApJ...450..559B} survey image of \obj, 
with rms of $0.135\rm\,mJy$, 
reveals a two-component morphology 
as shown in Figure~\ref{fig:opt_radio}. 
The core component is consistent in position with \obj\ in the optical 
and has an integrated flux density of $301.9\rm\,mJy$, 
while the secondary component is $\sim15\arcsec$ 
(i.e., projected distance of $120\rm\,kpc$ at rest frame) 
separated in north-west direction from the core 
and has an integrated flux density of $12.8\rm\,mJy$
and a side-lobe probability of only $P(S)$ = 0.025 
\citep[][]{2015ApJ...801...26H}.

In addition, 
Very Long Baseline Array (VLBA) observations of \obj\ at 5\,GHz revealed 
a core brightness temperature of $2.4\times10^{10}\rm\,K$ 
and a discernible jet feature with an angular extent less than 6 milliarcseconds, 
implying a jet on tens of parsec scale \citep[][]{2012ApJ...744..177L}. 
By using the Green Bank flux density $S_{4.85\rm\,GHz}=294\rm\,mJy$ \citep[][]{1992ApJS...79..331W} and the SDSS $g$-magnitude, 
we calculate the radio loudness of \obj\, 
as $\log R\approx3.7$ 
after the $k$-correction with radio spectral index $\alpha_{\rm rad}\sim-0.2$ \citep[][]{1992ApJS...79..331W} 
and optical spectral index $\alpha_{\rm opt}\approx-0.35$ obtained from the spectral fitting (Section~\ref{sec:spec_fitting}). 
Even considering radio variability, 
\obj\ is still an extremely radio-loud AGN, 
and one of the radio-loudest among known NLS1 galaxies. 

\subsection{Near-infrared and optical variability}
\label{sec:nir}

\obj\ was observed with the {\it Wide-field Infrared Survey Explorer} 
\citep[\wise;][]{2010AJ....140.1868W} in four bands 
$w1$, $w2$, $w3$ and $w4$ 
centred at 3.4, 4.6, 12 and 24\,\weimi, respectively. 
We employed the ALLWISE and NEOWISE database \citep[][]{2014ApJ...792...30M}. 
Due to the depletion of the satellite's cooling material \citep[][]{2014ApJ...792...30M},
we only constructed the Near-infrared (NIR) light curves of \obj\ in $w1$ and $w2$ 
which correspond to \obj\ emission at $\sim1.7$\,\weimi\ and 2.3\,\weimi\ at rest frame, respectively, 
We use $w1$ and $w2$ photometry data with S/N$>$10 and reduced $\chi^{2}$ lower than 2 based on profile fitting\footnote{The image processing and data quality details can be found in \url{http://wise2.ipac.caltech.edu/docs/release/allsky/expsup/}.}
in which case $\sim7\%$ of the data points are discarded.

\begin{figure}
	\centering
	\includegraphics[width=1.00\columnwidth]{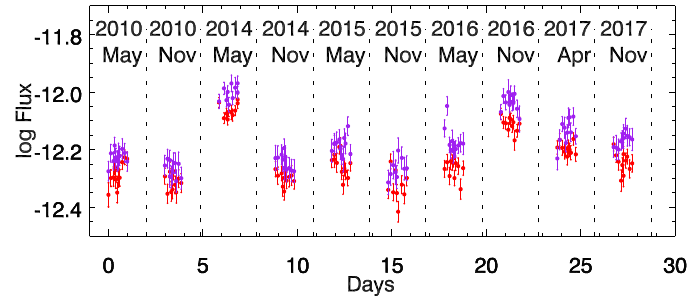}
	\caption{
	\wise\ light curves in the $w1$ (red) and $w2$ (purple) band in units of $\rm ergs\,s^{-1}cm^{-2}$ converted from profile-fit magnitudes \citep[][]{2010AJ....140.1868W}. 
	The MJDs of the light curves are offset for visual clarity.
	\label{fig:wise_lc}}
\end{figure}

The $w1$ and $w2$ light curves of \obj\ are displayed in Figure~\ref{fig:wise_lc}. 
Intra-day variability of $w1$ flux is detected with $p$-values of $P<0.5$ per\,cent using the $\chi^{2}$-test against the null hypothesis of no variation in 2015 November. 
The rapid variability restricts the size of the emitting region to be much smaller than the scale of the torus, 
implying a jet origin \citep[see also][]{2012ApJ...759L..31J, 2018rnls.confE..42G, 2019MNRAS.483.2362R}. 
On timescale of years, 
the NIR flux varied by nearly 0.4\,dex from May 2014 to November 2015. 
We also explore the color variation $(w1-w2)$. 
But no significant color variability has been found.

We have checked the archival photometry of the Catalina Real Time Transient Survey \citep[CRTS,][]{2009ApJ...696..870D} and the 
Panoramic Survey Telescope \& Rapid Response System \citep[Pan-STARRS,][]{2016arXiv161205560C} 
of \obj\ in the optical bands. 
CRTS has monitored \obj\ from May 2005 to January 2014. 
In spite of large uncertainties, 
the CRTS light curve reveals clear variability of \obj\ in the synthetic $V$-band, 
which brightened by $\sim0.5$ magnitude from 2006 to 2010 
and then became fainter again. 
Pan-STARRS monitored \obj\ in the $grizy$-bands between January 2010 and April 2014, 
and it also captured the slow decrease of the source's brightness after 2010 in all five bands.

\subsection{High-energy emission}
\label{sec:high_energy}

\subsubsection{X-ray emission}
The \rosat\ All-Sky Survey \citep[\rass;][]{1999A&A...349..389V} 
performed in 1990-1991 has covered the position of \obj. 
Following the method in \citet{2014ApJ...782...55Y}, 
we obtain a flux upper limit of $1.7\times10^{-13}$\,\flux\ in the band 0.1-2.4\,keV at the 99\% confidence level. 

\begin{figure}
	\centering
	\includegraphics[width=0.85\columnwidth]{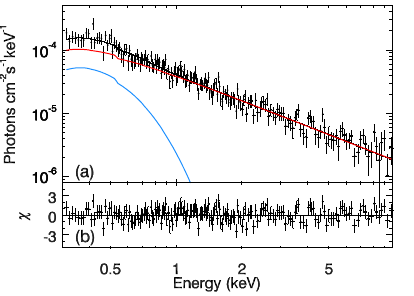}
	\caption{
	{\it(a)} The \xmm/PN unfolded spectrum and the best-fit model (black curve) consisting of power law (red) and black body (blue). 
	{\it(b)} Residuals of the power law plus black body model fitted to the spectrum. 
	\label{fig:pn}}
\end{figure}

We proposed an \xmm\ pointing observation of $\sim60$\,ks on \obj\ which was carried out on 2017 November 3 (ObsID 0800040101). 
The data are reduced with the \xmm\ {\sc Science Analysis System}\footnote{\url{https://www.cosmos.esa.int/web/xmm-newton/what-is-sas}}
({\sc version 15}) following standard procedures. 
The source position in the X-rays (R.A.=09:46:35.08, Dec=+10:17:06.04) is well consistent with that in SDSS image within 1 arcsec. 
The source events are extracted from a circle of 40\arcsec\ radius centred at source position for data from the PN CCD arrays {\citep{2001A&A...365L..18S}}, 
while the background events are extracted from a circle of the same radius in a source-free region nearby. 
The spectrum is binned to contain at least 25 counts per bin required for $\chi^{2}$ analysis. 
{\sc xspec v.12.9.0} {\citep[][]{1996ASPC..101...17A}} is used for spectral modeling over the 0.3$-$10\,keV energy range. 
We find that the spectrum can be well fitted ($\chi^{2}/\rm dof=193/183$) by a single power law and Galactic hydrogen absorption with column density of $N^{\rm Gal}_{\rm H}=2.45\times10^{20}\rm\,cm^{-2}$ \citep[][]{2005A&A...440..775K}. 
The fit is {slightly improved} ($\chi^{2}/\rm dof=172/181$) 
if we add a black body component, and we find a temperature of $kT_{\rm bb} = 0.23^{+0.07}_{-0.08}$ keV at rest frame 
(Figure~\ref{fig:pn}). 
The best-fit photon index is $\Gamma=1.35^{+0.08}_{-0.09}$,
which is relatively flat among the population of normal radio-quiet NLS1s \citep[e.g.][]{2010ApJS..187...64G}, 
but similar to other $\gamma$-ray emitting NLS1s \citep[$\Gamma<2$, e.g.][]{2009ApJ...707L.142A}. 
The total unabsorbed flux based on this model is $f_{0.3-10\rm\,keV}=4.04\times10^{-13}$\,\flux\ 
and $f_{0.1-2.4\rm\,keV}=1.38\times10^{-13}$\,\flux, 
with corresponding luminosities of 
$2.2\times10^{45}$\,\lum\ and 
$7.6\times10^{44}$\,\lum, respectively.

\subsubsection{$\gamma$-ray emission}

$\gamma$-ray emission from the direction of \obj\ was reported by \fermi. 
The source 4FGL~J0946.6+1016 was identified with the radio source TXS~0943+105 with association probability $>99.9\%$
\citep[][]{2019arXiv190210045T}. 
The significance of the $\gamma$-ray detection increases from $7\sigma$ in the LAT 1-year point source catalog 
\citep[1FGL,][]{2010ApJS..188..405A} 
to $46\sigma$ in the LAT 8-year point source catalog 
\citep[4FGL,][]{2019arXiv190210045T}. 
The flux (luminosity) between $100\rm\,MeV$ and $100\rm\,GeV$ also increases from 
$8.2\times10^{-12}$\,\flux\ ($4.5\times10^{46}$\,\lum) to $22.2\times10^{-12}$\,\flux\ ($1.2\times10^{47}$\,\lum), 
with a constant photon index of 2.4 assuming a power-law slope.

J0946+1017 underwent a giant $\gamma$-ray flare on July 2nd, 2014, 
with a daily average $\gamma$-ray flux about 25 times greater than the average flux reported in the 2FGL catalogue \citep{2014ATel.6292....1C}.

\section{Discussion and conclusions}
\label{sec:discussion}

J0946+1017 is the highest-redshift $\gamma$-ray emitting NLS1 known to date. 
It is one of the radio-loudest, and one of the few which show high-amplitude $\gamma$-ray flaring. Its Balmer decrement H$\alpha$/H$\beta \simeq 3$, 
implies little optical extinction. 
Its black hole mass,  
either derived from the Lorentzian fit 
($M_{\rm BH} = 1.6\times10^{8}\,M_{\odot}$) 
or Gaussian fit 
($M_{\rm BH} = 1.9\times10^{8}\,M_{\odot}$) 
based on the single-epoch scaling relation between 
Balmer line parameters and BH mass 
\citep[][]{2006ApJ...641..689V},
is at the lower end of BH masses in typical blazars, 
but at the high end of the NLS1 distribution. 
This result agrees well with the estimates of $M_{\rm BH}$ based on \mgii\ lines \citep[][]{2008ApJ...680..169S,2011ApJS..194...42R,2012ApJ...748...49S}.
Therefore \obj\ is an important cornerstone object bridging the two regimes. 
The monochromatic luminosity at 5100\,\AA\ is estimated to be $\lambda L_{5100}\approx1.2\times10^{45}$\,\lum\ from the H$\beta$ luminosity based on the relation given in \citet{2006ApJS..166..128Z}, 
as the observed flux at 5100\,\AA\ is likely contaminated by the beamed jet emission. 
With a bolometric correction of $k=9.8$ \citep[][]{2004MNRAS.352.1390M}, 
the Eddington ratio of \obj\ then is $\sim0.5$; 
typical of a NLS1.

While many of the radio-loud NLS1 galaxies have very compact radio emission in FIRST or in dedicated radio imaging 
\citep[e.g.][]{2006AJ....132..531K, 2015ApJS..221....3G, 2018A&A...614A..87B}, 
J0946+1017 is remarkable in showing widely extended radio emission in FIRST, on a (projected) spatial scale of 120 kpc. Several other NLS1 galaxies with very extended radio emission have been identified 
\citep[e.g.][]{2010ApJ...717.1243G, 2012ApJ...760...41D, 2015ApJ...800L...8R, 2018MNRAS.473.1554G, 2018ApJ...869..173R, 2018MNRAS.480.1796S}, however, 
only few of them exceed 100\,kpc 
(e.g., SDSS J103024.95+551622.7, \citealt{2018ApJ...869..173R}; 
SDSS J110006.07+442144.3, \citealt{2018MNRAS.473.1554G}; and
SDSS J094857.3+002225 with 395 kpc, \citealt{2006AJ....132..531K}).

\begin{figure}
	\centering
	\includegraphics[width=0.90\columnwidth]{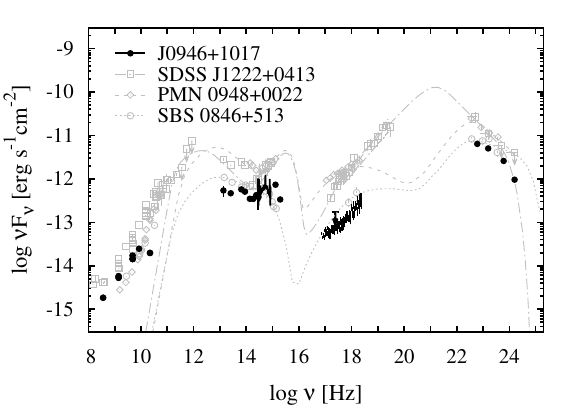}
	\caption{
	The SED of \obj\ (black) constructed with data from \wise, \boss, \rass, \xmm, the  \fermi\ 4-year point source catalog \citep[][]{2015ApJS..218...23A}, and data from the NASA/IPAC Infrared Science Archive and Extragalactic Database (NED). 
	The data are non-simultaneous and are corrected for Galactic
    reddening/absorption. 
	The SED of three other $\gamma$-ray detected NLS1s, 
	SDSS~J1222+0413 \citep[][]{2015MNRAS.454L..16Y}, 
	PMN~J0948+0022 \citep[][]{2012A&A...548A.106F} 
	and SBS~0846+513 \citep[][]{2012MNRAS.426..317D, 2015ApJ...798...43S} 
	and their models are also plotted for comparison. 
	\label{fig:sed}}
\end{figure}

The SED of J0946+1017 (Figure~\ref{fig:sed}) broadly resembles that of other $\gamma$-ray NLS1s (with three peaks
in the radio, ultraviolet and $\gamma$-ray bands, respectively, \citealt{2015MNRAS.454L..16Y}), 
which have been 
well fitted by leptonic jet models, which require external Compton scattering in the high-energy regime, with seed photons from the accretion disk, broad-line region, and/or torus. 
The overall flatness of the X-ray spectrum implies the dominance of jet emission, but the XMM spectrum can be well fit by a two-component model including thermal emission at low energies. 
The IR variability on timescales of months to days implies that the emission is not from the dusty torus, but rather related to the jet. 
Multi-band monitoring of future high-amplitude flaring of J0946+1017 will provide us with important probes of the jet physics in the radio-loudest AGN.

\section*{Acknowledgements}

We thank the anonymous referee for the very helpful comments and A. Lobanov for useful remarks. 
SY thanks A. Y. Yang for her help with plotting Fugire 2. 
SY acknowledges support by the KIAA-CAS Fellowship, which is jointly supported by Peking University and the Chinese Academy of Sciences. 
WJL acknowledges supports from the Natural Science Foundation of China (NSFC) grant No. 11703079 and the "Light of West China" Programme of Chinese Academy of Sciences (CAS). 
WY thanks the support from the National Science Foundation of China (NSFC-11703076) and the West Light Foundation of the Chinese Academy of Sciences (Y6XB016001). 
XBW thanks the supports by the Ministry of Science and Technology of China under grant 2016YFA0400703, 
the NSFC grants No.11721303 and 11533001.
This research made use of data taken from \sdss-\boss, FIRST, \wise, CRTS, Pan-STARRS, and \xmm. 
This research has made use of data taken from TripleSpec on P200 telescope through the Telescope Access Program (TAP). 
This research has made use of the NASA/IPAC Infrared Science Archive and Extragalactic Database (NED), which are operated by the Jet Propulsion Laboratory, California Institute of Technology, under contract with the National Aeronautics and Space Administration.

\bibliographystyle{mnras}
\bibliography{references}

\bsp	
\label{lastpage}
\end{document}